\documentclass[american]{article}
\usepackage[T1]{fontenc}
\usepackage[latin9]{inputenc}
\usepackage{color}
\usepackage{amsmath}
\usepackage{amsthm}
\usepackage{amssymb}
\usepackage{geometry}
\usepackage{esint}

\makeatletter

\newcommand{\noun}[1]{\textsc{#1}}

\numberwithin{equation}{section}
\numberwithin{figure}{section}

\makeatother

\usepackage{babel}
\begin{document}
\title{Origin of the quantum operator formalism and its connection with linear
response theory}
\author{Ana María Cetto\thanks{Corresponding author. Email: ana@fisica.unam.mx}
$\:$and Luis de la Peña}
\maketitle
\begin{center}
Instituto de Física, Universidad Nacional Autónoma de México, Mexico
\par\end{center}
\begin{abstract}
Linear response theory is concerned with the way in which a physical
system reacts to a small change in the applied forces. Here we show
that quantum mechanics in the Heisenberg representation can be understood
as a linear response theory. To this effect, we first address the
question of the physical origin of the quantum operator formalism
by considering the interaction of a bound electron with the radiation
field, including the zero-point component, following the approach
of stochastic electrodynamics. Once the electron has reached a stationary
state, it responds linearly and resonantly to a set of modes of the
driving radiation field. Such a response can lead the system to a
new stationary state. Identifying a one-to-one relationship between
the response variables and the corresponding operators, results in
the (\emph{x,p}) commutator as the Poisson bracket of these variables
with respect to the driving field amplitudes. To account for the order
of the response variables, which is reflected in the non-commutativity
of the operators, we introduce the concept of ordered covariance.
The results obtained allow to establish a natural contact with linear
response theory at the fundamental quantum level.
\end{abstract}

\section{Introduction}

In a chapter on linear response functions, E. Pavarini \cite{Pava}
writes ``All we know about a physical system stems either from its
effects on other physical systems or from its response to external
forces.'' This resonates with Heisenberg's insightful work of 1925,
which led to the matrix formulation of quantum mechanics: the inside
of atoms is unknown to us, but he had the spectroscopic data that
tell us how the atoms emit or absorb radiation, which he aptly codified
using a non-commutative algebra. 

Heisenberg later took things to an extreme by claiming that what we
observe is all that exists. A hundred years on, we know a great deal
more about atoms, but matrix mechanics and its central element, the
quantum commutator, still have the flavor of a postulate, a given
that does not permit deeper investigation. The obscurity surrounding
their physical meaning has far-reaching consequences for the understanding
of quantum phenomena, which are still the subject of much debate and
confusion.

It is in this context that we have undertaken an in-depth study of
the emergence of quantum operators, the results of which are documented
in recent papers (\cite{foop22,foop24} and references therein). The
leitmotif of this research has been precisely the coupling of matter
(in particular atomic electrons) to the radiation field, with the
purpose of deriving the effects of this coupling on both sides of
the system. In accordance with stochastic electrodynamics (\noun{sed}),
a causal theory for quantum mechanics constructed from first physical
principles, the radiation field includes its vacuum component, the
zero-point field (\noun{zpf}). Electrons, like all electromagnetic
matter, constantly interact with this field; and the central effect
of this interaction is shown to be indeed the quantization of matter--as
well as of the field. 

As a charged particle, the electron radiates constantly. This problem
has been known since the beginning of atomic theory and remained unsolved
for decades until the pioneering work in \noun{sed} put forward a
solution by invoking the \noun{zpf} (\cite{Mars63,Dice} and references
therein)---in line with a suggestion made by Nernst way back in 1916
\cite{Nernst}. We can now ascertain that after a fluctuating dissipative,
and hence irreversible, process, the atomic electron reaches a stationary
state thanks to the inevitable presence of the \noun{zpf}, which systematically
replaces the energy lost by radiation reaction \cite{TEQ}. When the
electron is in a stationary state, it responds resonantly to certain
modes of the field, those that can bring it to a different stationary
state.

We should recall that the general purpose of linear response theory
is to determine the response of the system to the stimulus of an external
force, identified as the ``driving force'' \cite{Kubo,Kubobook}.
By studying the response of the system at different frequencies, important
information is obtained about the properties of the system itself.
Linear response theory has been successfully used in quantum statistical
mechanics to derive the optical and electromagnetic properties of
materials exposed to fields that are not too strong, but stronger
than the \noun{zpf}. Here, by contrast, we use classical \noun{lrt}
to confirm that, from the perspective of \noun{sed}, the quantum matrix
formalism is indeed a linear response description of the (atomic)
matter-field interaction.

In preparation for the central part of the paper, in Section 2 we
present the basic relevant elements of \noun{sed} that are required
for a correct understanding of the subsequent discussion, with a focus
on the dynamical behavior of a charged particle subject to a binding
force in addition to the radiation field. In Section 3, we show that
the irreversible process leading to the establishment of the quantum
regime involves a qualitative change in the nature of the variables
used for the description, which now represent the response to the
driving radiation field. Section 4 considers the \noun{sed} system
from the perspective of \noun{lrt }and thus serves as a bridge to
Section 5, where the connection between the \noun{sed} response functions
and the corresponding quantum operators is presented. In Section 6,
a one-to-one relation is established between the \noun{sed} ordered
covariance and the quantum anticommutator, which complements the previously
established one-to-one relation between the \noun{sed} canonical Poisson
bracket and the canonical quantum commutator. The paper concludes
with a brief discussion of the main findings.

\section{The stochastic process underlying quantum mechanics}

We recall that stochastic electrodynamics (\noun{sed}) has been developed
on the basis of the existence of the \noun{zpf} as a real, fluctuating
Maxwellian field in permanent interaction with matter, with the aim
of providing a causal, physical explanation for quantum mechanics.
Here we present the basic relevant elements of \noun{sed} that are
required for a correct understanding of the following sections.

\subsection{The particle-field system in the Markov approximation}

The usual starting point of \noun{sed} is the equation of motion for
a charged particle---typically an atomic electron---subject to a
conservative binding force in addition to the background radiation
field. This field includes by default the random zero-point radiation
field (\noun{zpf}) with an energy $\hbar\omega/2$ per mode \cite{TEQ},
which corresponds to a spectral energy density 
\begin{equation}
\rho_{0}(\omega)=\frac{\hbar\omega^{3}}{2\pi^{2}c^{3}},\label{eq:10}
\end{equation}
and hence an autocorrelation of the electric component given by 
\begin{equation}
\langle E_{i}(t^{\prime})E_{j}(t)\rangle_{0}=\varphi(t'-t)\delta_{ij}=\frac{2\pi}{3}\delta_{ij}\int\rho_{0}(\omega)e^{i\omega(t^{\prime}-t)}d\omega.\label{23.11-1}
\end{equation}
We will in general limit the discussion for the sake of simplicity
to one dimension. Then the Langevin equation of motion for the electron
is 
\begin{equation}
m\ddot{x}-f(x)-m\tau\dddot{x}=eE(t),\label{2}
\end{equation}
where $\tau=2e^{2}/3mc^{3}$ and the electric field is taken in the
long-wavelength (dipole) approximation. Different procedures, well
known from the theory of stochastic processes \cite{vanK,Pap91},
lead from Eq. (\ref{2}) to a generalized Fokker-Planck equation for
the particle phase-space probability density, which is an integro-differential
equation---or equivalently, a differential equation with an infinite
number of time-dependent terms---that is impossible to solve exactly.
However, some relevant observations can be drawn from it.

Initially, when particle and field get connected, the system is far
from equilibrium. In this regime, the main effect of the \textsc{zpf}
on the particle is due to the modes of very high frequency, close
to $10^{21}\mathrm{s}^{-1}$, which produce violent accelerations
and randomize the motion. Eventually, the interplay between the electric
field force and radiation reaction drives the system irreversibly
into equilibrium; in this regime, the initial conditions have become
irrelevant and the Markovian approximation applies. The time averaging
implicit in the Markov description prevents the quantum description
from capturing the finer details of the motion. 

We can estimate the time resolution of the Markov description by resorting
to stochastic quantum mechanics, \noun{sqm}, the phenomenological
theory that exhibits quantum mechanics as a Markov stochastic process.
As shown in Refs. \cite{Dice,Nel66,Nel12}, \noun{sqm} requires the
introduction of a finite time interval $\Delta t$; thus, any phenomenon
or detail shorter in time than $\Delta t$ remains outside the possibilities
of (non-relativistic) quantum theory. In \noun{sqm}, the diffusion
coefficient $D,$which is related to the variance of $x$ by $\overline{\left(\Delta x\right)^{2}}=2D\Delta t$,
is phenomenologically adscribed the value $D=\hslash/2m$. Combining
this with the above equation gives
\begin{equation}
\overline{\left(\Delta x\right)^{2}}=\left(\hslash/m\right)\Delta t.\label{Q12}
\end{equation}
To estimate $\Delta t$ we assign a minimum characteristic value to
the mean velocity $\bar{v}=\sqrt{\overline{\left(\Delta x\right)^{2}}}/\Delta t$,
and consider an orbiting atomic electron; its kinetic energy is of
the order $\alpha^{2}mc^{2}$ , where $\alpha$ is the fine structure
constant; hence 
\begin{equation}
\overline{\left(\Delta x\right)^{2}}/\left(\Delta t\right)^{2}=\frac{\hslash}{m\Delta t}\simeq\alpha^{2}c^{2}.\label{Q14}
\end{equation}
It follows that 
\begin{equation}
\Delta t\simeq\frac{\hslash}{m\alpha^{2}c^{2}}=\frac{1}{\alpha^{2}\omega_{C}}=\frac{1}{2\pi\alpha^{2}}\tau_{C},\label{Q16}
\end{equation}
where $\tau_{C}$ is the Compton time of the electron. With the above
estimates, we get $\Delta t\sim10^{-17}\,s$, a time interval that
is still beyond any experimental limit.

Note that with a dispersion of $p$ of the order of $m\bar{v}$, the
above equations give $\overline{\left(\Delta x\right)^{2}}\,\overline{\left(\Delta p\right)^{2}}\simeq\hbar^{2}$,
implying that the limited time resolution results in a limited resolution
in $x$ and $p$.

In the Markov approximation, describing the slow dynamics, the generalized
Fokker-Planck equation is simplified by retaining terms up to second
order. The ensuing Fokker-Planck equation ($x$, $p$ are the Cartesian
coordinates and corresponding momenta),
\begin{equation}
\frac{\partial Q}{\partial t}+\frac{1}{m}\frac{\partial}{\partial x_{}}p_{}Q+\frac{\partial}{\partial p_{}}{}_{i}Q=-m\tau\frac{\partial}{\partial p_{}}\dddot{x}_{}Q+D^{px}\frac{\partial^{2}Q}{\partial p_{}\partial x_{}}+D_{ij}^{pp}\frac{\partial^{2}Q}{\partial p_{}\partial p_{}},\label{3.C28-1}
\end{equation}
contains remnants of the memory build-up, expressed through the diffusion
tensors 
\begin{equation}
D_{i}^{px}(t)=e\left\langle x_{}E_{}\right\rangle _{0}=e^{2}\int_{-\infty}^{t}ds\left.\frac{\partial x_{}(t)}{\partial p_{}(s)}\right|_{x^{(0)}}\left\langle E(s)E_{}(t)\right\rangle _{0},\label{3.C10a-1}
\end{equation}
\begin{equation}
D^{pp}(t)=e\left\langle p_{}E_{}\right\rangle _{0}=e^{2}\int_{-\infty}^{t}ds\left.\frac{\partial p_{}(t)}{\partial p_{}(s)}\right|_{x^{(0)}}\left\langle E_{}(s)E(t)\right\rangle _{0}.\label{3.C10b-1}
\end{equation}
 Equation (\ref{3.C28-1}) provides key statistical information about
the evolution of the system in the quantum regime. Multiplying it
from the left by a dynamical function $\mathcal{G}(x,p,t)$ and taking
the average one gets
\begin{equation}
\frac{d}{dt}\left\langle \mathcal{G}\right\rangle =\left\langle \frac{d\mathcal{G}}{dt}\right\rangle _{\text{nr}}+m\tau\left\langle \dddot{x}\frac{\partial\mathcal{G}}{\partial p}\right\rangle -e^{2}\left\langle \frac{\partial\mathcal{G}}{\partial p}\mathcal{\hat{D}}\right\rangle ,\label{m11a-1}
\end{equation}
where
\begin{equation}
\left\langle \frac{d\mathcal{G}}{dt}\right\rangle _{\text{nr}}=\left\langle \frac{\partial\mathcal{G}}{\partial t}\right\rangle +\left\langle \dot{x}\frac{\partial\mathcal{G}}{\partial x}+f\frac{\partial\mathcal{G}}{\partial p}\right\rangle \label{m11c-1}
\end{equation}
corresponds to the (Liouvillian) non-radiative contribution to $\left\langle d\mathcal{G}/dt\right\rangle $.
This equation describes the evolution of the mean value of $\mathcal{G}$
in line with quantum mechanics, i.e. in the radiationless approximation.
The remaining terms in Eq. (\ref{m11a-1}), with the diffusion operator
$\mathcal{\hat{D}}$ given by
\begin{equation}
e^{2}\mathcal{\hat{D}}=D^{px}\frac{\partial}{\partial x}+D^{pp}\frac{\partial}{\partial p},\label{m11d-1-1}
\end{equation}
represent the radiative corrections. When $\mathcal{G}=\xi(x,p)$
represents an integral of motion, (\ref{m11c-1}) gives 
\begin{equation}
\left\langle \dot{x}\frac{\partial\mathcal{\xi}}{\partial x_{}}+f\frac{\partial\mathcal{\xi}}{\partial p}\right\rangle =0\label{m22}
\end{equation}
and (\ref{m11a-1}) reduces to
\begin{equation}
\frac{d}{dt}\left\langle \xi\right\rangle =m\tau\left\langle \dddot{x}\frac{\partial\xi}{\partial p}\right\rangle -e^{2}\left\langle \frac{\partial\xi}{\partial p_{}}\mathcal{\hat{D}}\right\rangle ,\label{m24-2}
\end{equation}
which shows that only the radiative terms contribute to the evolution
of $\left\langle \xi\right\rangle $. The system reaches a state of
equilibrium when
\begin{equation}
m\tau\left\langle \dddot{x}\frac{\partial\xi}{\partial p_{}}\right\rangle =e^{2}\left\langle \frac{\partial\xi}{\partial p}\mathcal{\hat{D}}_{}\right\rangle .\label{m23}
\end{equation}

This is a general \emph{fluctuation-dissipation relation}: the left
side represents the average loss of $\xi$ per unit time due to the
radiation reaction and the right term represents the average exchange
of $\xi$ per unit time with the background field. In particular,
when $\xi$ is the Hamiltonian, (\ref{m23}) implies that the system
has reached an energy eigenstate. \emph{It is precisely the combined
effect of radiation (dissipation) and diffusion that allows the system
to achieve the equilibrium necessary to remain in a stationary quantum
state }\cite{TEQ}\emph{.}

Equation (\ref{m24-2}) has been successfully used to obtain the radiative
corrections which otherwise require recourse to (nonrelativistic)
quantum electrodynamics \cite{Mil}, in particular the correct formulae
for the atomic lifetimes and the Lamb shift \cite{TEQ}.

\section{Description of the quantum regime. The new kinematics}

The attainment of the quantum regime entails an important change of
nature of the particle dynamical variables used for the description,
i.e. of the kinematics. To analyze this change we start from the Poisson
bracket of the particle's canonical variables at time $t$, which
in three-dimensional notation reads 
\begin{equation}
\left\{ x_{j},p_{i}\right\} _{xp}=\sum_{k}\left(\frac{\partial x_{j}}{\partial x_{k}}\frac{\partial p_{i}}{\partial p_{k}}-\frac{\partial p_{i}}{\partial x_{k}}\frac{\partial x_{j}}{\partial p_{k}}\right)\label{eq:nre}
\end{equation}
and must sutisfy 
\begin{equation}
\left\{ x_{j},p_{i}\right\} _{xp}=\delta_{ij},\label{K4-1}
\end{equation}
where $i,j,k=1,2,3$, and the variables and the derivatives are taken
at the same time $t$.

The \emph{full} set of canonical variables at any time comprises both
those of the particle, $\left\{ x_{i};p_{i}\right\} $, and those
of the field modes, $\left\{ \mathrm{q}_{\alpha};\mathrm{p}_{\alpha}\right\} $
(roman typography is used for the field's canonical variables to distinguish
them from those of the particle, a semicolon is used to distinguish
a set of variables from a Poisson bracket. and a discrete set $\left\{ \alpha\right\} $
of field modes is considered for reasons that will become clear later),
i.e.
\begin{equation}
\left\{ q;p\right\} =\left\{ x_{i},\mathrm{q}_{\alpha};p_{i},\mathrm{p}_{\alpha}\right\} .\label{K6}
\end{equation}
At the initial time $t_{o}$, when particle and field begun to interact,
the canonical variables were
\begin{equation}
\left\{ q_{o};p_{o}\right\} =\left\{ x_{io},\mathrm{q}_{\alpha o};p_{io},\mathrm{p}_{\alpha o}\right\} .\label{K8-1}
\end{equation}
Since the whole system is Hamiltonian, the variables at times $t_{0}$
and $t$ are related by a canonical transformation, and the particle's
Poisson bracket at time $t$ can be taken with respect to either set
of variables,
\begin{equation}
\left\{ x,p\right\} _{xp}=\left\{ x,p\right\} _{x_{o}p_{o}}+\left\{ x,p\right\} _{\mathrm{q}_{\alpha o}\mathrm{p}_{\alpha o}}.\label{K10}
\end{equation}
Therefore, according to Eq. (\ref{K4-1}),
\begin{equation}
\left\{ x_{i}(t),p_{j}(t)\right\} _{x_{o}p_{o}}+\left\{ x_{i}(t),p_{j}(t)\right\} _{\mathrm{q}_{\alpha o}\mathrm{p}_{\alpha o}}=\delta_{ij}.\label{K14-1}
\end{equation}
Since in the quantum regime the particle has lost track of its initial
conditions, the first term in Eq. (\ref{K14-1}) vanishes,
\begin{equation}
\left\{ x_{i}(t),p_{j}(t)\right\} _{q_{o}p_{o}\:}\rightarrow\left\{ x_{i}(t),p_{j}(t)\right\} _{\mathrm{q}_{\alpha o}\mathrm{p}_{\alpha o}},\label{K16-1}
\end{equation}
and the particle's Poisson bracket becomes defined by the canonical
variables of the field modes $\alpha$ with which it interacts, 
\begin{equation}
\left\{ x_{i}(t),p_{j}(t)\right\} _{\mathrm{q}_{\alpha o}\mathrm{p}_{\alpha o}}=\delta_{ij}.\label{K18-1}
\end{equation}
Taking into account that a field mode of frequency $\omega_{\alpha}$
carries with it an energy $\hbar\omega_{\alpha}$, we introduce the
normal field amplitudes $a_{\alpha},a_{\alpha}^{*}$, related to $\mathrm{q}_{\alpha o},\mathrm{p}_{\alpha o}$
by the transformation 
\begin{equation}
\omega_{\alpha}\mathrm{q}_{\alpha o}=\sqrt{\hbar\omega_{\alpha}/2}(a_{\alpha}+a_{\alpha}^{*}),\;\mathrm{p}_{\alpha o}=-i\sqrt{\hbar\omega_{\alpha}/2}(a_{\alpha}-a_{\alpha}^{*}),\label{K20}
\end{equation}
 The complex field amplitudes are normalized to unity, i.e. $a_{\alpha}=e^{i\phi_{\alpha}},a_{\alpha}^{*}=e^{-i\phi_{\alpha}},$
where $\phi_{\alpha}$ are statistically independent random phases.
The Poisson bracket of two functions $f,g$ with respect to $a_{\alpha},a_{\alpha}^{*}$
is thus 
\begin{equation}
\left\{ f,g\right\} _{aa^{*}}=\sum_{\alpha}\left(\frac{\partial f}{\partial a}_{\alpha}\frac{\partial g}{\partial a_{\alpha}^{*}}-\frac{\partial g}{\partial a}_{\alpha}\frac{\partial f}{\partial a_{\alpha}^{*}}\right)=i\hbar\left\{ f,g\right\} _{q_{\alpha o}p_{\alpha o}}.\label{K26}
\end{equation}
Applied to the particle's canonical variables, (\ref{K26}) reads
\begin{equation}
i\hbar\left\{ x_{i},p_{j}\right\} _{\mathrm{q}_{\alpha o}\mathrm{p}_{\alpha o}}=\left\{ x_{i},p_{j}\right\} _{aa^{*}},\label{K24-3}
\end{equation}
and therefore, according to Eq. (\ref{K18-1}), for the particle in
a stationary state we get
\begin{equation}
\left\{ x_{i},p_{j}\right\} _{aa^{*}}=i\hbar\delta_{ij}.\label{K28-1}
\end{equation}
This result indicates that the relation between the particle variables
$x_{i}$ and $p_{j}$\emph{ becomes determined by their functional
dependence on the normal}\emph{\noun{ }}\emph{field variables $\{a_{\alpha};a_{\alpha}^{*}$\},
}with the scale given by Planck's constant. The (classical) symplectic
structure is preserved, but the meaning of the quantities $x_{i},p_{j}$
has changed.

\section{The perspective of linear response theory (\noun{lrt})}

If the system, already in a stationary state in the quantum regime,
is subjected to an external field or force that varies with time,
its response to external stimuli will be governed by the dynamics
it has acquired in that regime. The \noun{zpf} has fulfilled its primary
function of bringing the system into the quantum regime, but it will
continue to be present, so it can also stimulate a response from the
particle. The response of the particle to the stimuli of the radiative
field (whether the \noun{zpf} alone or in combination with an external
field) is precisely what takes it from one stationary state to another,
i.e. what makes it undergo a radiative transition.

When quantum mechanics considers an energy eigenstate $n$, it works
with a particular stationary distribution: that of the systems that
possess exactly that energy. Take for example an ensemble of harmonic
oscillators: in the non-radiative approximation (which corresponds
to Schrödinger's quantum mechanics), all the ensemble's oscillators
are in energetic equilibrium with the modes of the field with which
they resonate, as we will see below; the only thing that distinguishes
the elements of the ensemble is the phase. When we talk about the
phase of the field modes, in reality the relevant quantity is the
phase with which they are coupled to the harmonic oscillator (or the
oscillator is coupled to them), and this differs from one element
of the ensemble to another. 

The harmonic oscillator that has reached a state of stationary of
motion resonates with the field modes of frequency equal to its natural
frequency $\omega_{0}$. In the general case of a non-linear binding
force, the particle resonates at more than one field frequency; this
can be a finite or infinite number of frequencies, depending on the
case. Since the phases corresponding to the modes of different frequencies
are statistically independent, the appearance of more than one frequency
does not introduce dispersion in the energy; the energies associated
with the resonances remain fixed.

In order to look at the problem from the perspective of linear response
theory (\noun{lrt}) we have to bear in mind that the force $eE(t)$
on the right-hand side of Eq. (\ref{2}) is given, it is not dynamical.
Classical \noun{lrt} allows us to determine the reaction of the system
to an external driving force when its intensity is low; under this
condition the change of $x_{n}(t)$ is given by \cite{Wio}
\begin{equation}
x_{n}(t)=\frac{e}{m}\intop_{-\infty}^{+\infty}dt'\chi_{n}(t-t')E(t').\label{A4}
\end{equation}
The response of the system to a sinusoidal force, identified as the
electric susceptibility, is given by the Fourier transform of the
response function $\chi(t)$,
\begin{equation}
\widetilde{x}(\omega)=\frac{e}{m}\widetilde{\chi}(\omega)\widetilde{E}(\omega).\label{A6}
\end{equation}
In the case of the \noun{sed} harmonic oscillator, the solution of
Eq. (\ref{2}) gives 
\[
\widetilde{\chi}_{kn}(\omega)=\frac{1}{\omega_{kn}^{2}-\omega^{2}-i\tau\omega^{3}},
\]
with $\omega_{kn}=\pm\omega_{0}$. When there is more than one response
frequency, the susceptibility is given by the sum of all responses,
$\widetilde{\chi_{n}}(\omega)=\sum_{k}\widetilde{\chi}_{kn}(\omega)$,
with
\begin{equation}
\widetilde{\chi}_{kn}(\omega)=\frac{1}{\omega_{kn}^{2}-\omega^{2}-i\tau\omega^{3}},\label{A12}
\end{equation}
and the response function is therefore given by
\begin{equation}
\chi_{n}(t)=\frac{1}{2\pi}\intop_{-\infty}^{+\infty}d\omega e^{-i\omega t}\sum_{k}\widetilde{\chi}_{kn}(\omega),\label{51}
\end{equation}
where $\mathrm{Re}\widetilde{\chi}_{kn}(\omega)$ represents a reactive
term and $\mathrm{Im}\widetilde{\chi}_{kn}(\omega)$ a dissipative
or absorptive term. These are related by the Kramers-Kronig formula
\begin{equation}
\widetilde{\chi}_{kn}(\omega)=\intop_{-\infty}^{+\infty}\frac{d\omega'}{\pi}\frac{\mathrm{Im}\widetilde{\chi}_{kn}(\omega')}{\omega'-\omega_{kn}-i\tau\omega'{}^{2}},\label{A14}
\end{equation}
indicating that the higher the reactive response to a given $\omega_{kn}$,
the more intense is the absorption or emission at $\omega\sim\omega_{kn}$.

From the above equations it follows that the Fourier transform of
the spectrum $S_{x}(\omega)$ is \cite{Wio,ES2022} 
\begin{equation}
\intop_{0}^{+\infty}d\omega S_{x}(\omega)e^{-i\omega(t'-t)}=\left\langle x_{n}(t)x_{n}(t')\right\rangle +\frac{1}{2}\left[x_{n}(t),x_{n}(t')\right],\label{A14b}
\end{equation}
where $\left\langle \cdot\right\rangle $ indicates average value
over the random phases and $\left[\cdot\right]$ indicates Poisson
bracket. Calculation of the first term gives \emph{for every single}
$\omega_{kn}$, with the \noun{zpf} spectrum given by Eq. (\ref{eq:10}),
\begin{equation}
\left\langle x_{nk}(t)x_{kn}(t)\right\rangle =\frac{\hbar\tau}{\pi m}\intop_{0}^{+\infty}d\omega\frac{\omega^{3}}{(\omega_{kn}^{2}-\omega^{2})^{2}+\tau^{2}\omega^{6}}.\label{A14c}
\end{equation}

\noindent For atomic frequencies, $1/\tau\omega_{kn}\approx10^{8}-10^{9}$
and the integrand can be approximated by a delta funtion; in this
approximation the integral gives $\left\langle x_{nk}x_{kn}\right\rangle =\hbar/\left\{ m\left|\omega_{kn}\right|\right\} $.
This confirms that the particle resonates very sharply to the frequencies
$\omega_{kn}$.

\section{Origin of the quantum operators}

The previous discussion allows us to identify the field modes $\alpha$
of Section 3 as those to which the particle responds resonantly, i.e.
$\left\{ \alpha\right\} \Rightarrow\left\{ nk\right\} $, and to write
instead of (\ref{K28-1}) (in one-dimensional notation)
\begin{equation}
\left\{ x_{},p\right\} _{a_{nk}a_{nk}^{*}}=i\hbar,\label{Oo2}
\end{equation}
for the Poisson bracket of $x,p$ with respect to the corresponding
set of field variables $\left\{ a_{nk};a_{nk}^{*}\right\} $ that
can take the particle from state $n$ to any other state $k$. The
constant value $i\hbar$ of this bilinear form implies that $x_{n}$
and $p_{n}=m\dot{x}_{n}$ are linear functions of the field variables,
so that they can be expressed in the general form
\[
x_{n}(t)-x_{nn}=\sum_{k}x_{nk}a_{nk}e^{-i\omega_{kn}t}\mathrm{+c.c.,\:\;}
\]
\begin{equation}
p_{n}(t)-p_{nn}=m\dot{x}_{n}\sum_{k}p_{nk}a_{nk}e^{-i\omega_{kn}t}\mathrm{+c.c.,}\label{T4}
\end{equation}
where $a_{nk}$ connects state $n$ with state $k$, and $x_{nk}$
is the response coefficient. 

\noindent Introduced in Eq. (\ref{Oo2}), this gives
\begin{equation}
\left\{ x,p\right\} _{nn}=2im\sum_{k}\omega_{kn}\left|x_{nk}\right|^{2}=i\hbar,\label{T5}
\end{equation}
\begin{equation}
\mathrm{whence}\quad\quad\sum_{k}\omega_{kn}\left|x_{nk}\right|^{2}=\hbar/2m.\label{T6}
\end{equation}
More generally, since $a_{nk},a_{n'k}$ have independent random phases
for $n'\neq n$, 
\begin{equation}
\left\{ x,p\right\} _{nn'}=i\hbar\delta_{nn'}.\label{T8}
\end{equation}
The $x_{nk}$, $a_{nk}$ refer to the transition $n\rightarrow k$
; $x_{kn}$, $a_{kn}$ refer to the inverse transition, with $\omega_{nk}=-\omega_{kn}$;
therefore, $x_{nk}^{*}(\omega_{nk})=x_{kn}(\omega_{kn}),\ p_{nk}^{*}(\omega_{nk})=p_{kn}(\omega_{kn}),\ a_{nk}^{*}(\omega_{nk})=a_{kn}(\omega_{kn}),$
whence from (\ref{T8}),
\begin{equation}
{\displaystyle \sum_{k}}\left(x_{nk}p_{kn'}-p_{n'k}x_{kn}\right)=i\hbar\delta_{nn'}.\label{T12}
\end{equation}

\noindent Identifying $x_{nk}$ and $p_{nk}$ as the elements of two
matrices $\hat{x}$ and $\hat{p}$, Eq. (\ref{T12}) becomes $\left[\hat{x},\hat{p}\right]_{nn'}=i\hbar\delta_{nn'}$,
i.e. the matrix formula for \textcolor{black}{
\begin{equation}
{\color{red}\mathinner{\normalcolor \left[\hat{x},\hat{p}\right]}{\normalcolor =i\hbar}\mathpunct{\normalcolor ,}}\label{T16}
\end{equation}
which shows that} \emph{the quantum canonical commutator is the Poisson
bracket of the system response variables $x,p$ with respect to the
normal field variables} $\left\{ a;a^{*}\right\} $. This provides
a causal explanation for the appearance of operators in quantum mechanics.

Equation (\ref{T16}) is used as the basis for the general transformation
of dynamical variables into operators. Once $x$ and $p$ become operators,
all dynamical variables $G(x,p;t)$ become operators $\widehat{G}(\widehat{x},\widehat{p};t)$
that act on the states, which are represented by column matrices.
Therefore, when applying Eq. (\ref{m11a-1}) and the following to
study the dynamics in the quantum regime, one must express them in
terms of operators and replace the averages by expectation values.
In particular, this means that the Liouvillian (radiationless) equation
(\ref{m11c-1}), which can be written as the average of Hamilton's
equation (with $H=(\boldsymbol{p}^{2}/2m)+V$),
\begin{equation}
\left\langle \dot{G}\right\rangle =\left\langle \frac{\partial\mathcal{G}}{\partial t}\right\rangle +\left\langle \left\{ G,H\right\} \right\rangle ,\label{m26-1}
\end{equation}
takes the form 
\begin{equation}
\left\langle \dot{\hat{G}}\right\rangle =\left\langle \frac{\partial\mathcal{\hat{G}}}{\partial t}\right\rangle +i\hbar\left\langle \left[\hat{G},\hat{H}\right]\right\rangle ,\label{m28-1}
\end{equation}
where the angular bracket is to be interpreted as the expectation
value of the Heisenberg equation. Experience shows that this replacement
is usually straightforward, although occasionally some additional
analysis is required to take account of the order of the (non-commuting)
operators. 

\section{Ordered covariances and quantum expectation values}

In quantum mechanics the expectation values of dynamical quantities
are calculated in terms of matrix elements. When non-commuting operators
are involved, there can be more than one expectation value, depending
on the order of the operators, e.g. 
\begin{equation}
\left\langle \hat{x}\hat{p}\right\rangle _{n}-x_{nn}p_{nn}=im\sum_{k}\omega_{kn}\left|x_{kn}\right|^{2},\label{eq:065}
\end{equation}
\begin{equation}
\left\langle \hat{p}\hat{x}\right\rangle _{n}-x_{nn}p_{nn}=-im\sum_{k}\omega_{kn}\left|x_{kn}\right|^{2}.\label{eq:066}
\end{equation}

Let us look at it from the \noun{sed} perspective, i.e. without resorting
to the operator language. We consider a stationary state $n$; the
$x_{kn},p_{kn}$ are the coefficients of the response function involved
in the transition $n\leftrightarrow k$. This allows us to calculate
quantities such as 
\begin{equation}
C_{n}(xp)=\left\langle \left(x_{n}^{*}(t)-x_{nn}\right)\left(p_{n}(t)-p_{nn}\right)\right\rangle ,\label{eq:067}
\end{equation}
the ordered covariance of the response function $\left(p_{n}(t)-p_{nn}\right)$
that can take the particle from state $n$ to any state $k$, followed
(on the left, to respect the quantum convention) by the response function
$\left(x_{n}^{*}(t)-x_{nn}\right)$ that brings the particle from
state $k$ back to state $n$. The conjugated left term compensates
the effect of the right term. 

If instead $x_{n}(t)$ takes the particle from $n$ to any $k$, and
$p_{n}(t)$ brings it back to $n$, we get the covariance $C_{n}(px)$.
Thus, 
\begin{equation}
C_{n}(xp)=\left\langle x^{*}(t)p(t)\right\rangle _{nn}-x_{nn}p_{nn},\quad C_{n}(px)=\left\langle p^{*}(t)x(t)\right\rangle _{nn}-x_{nn}p_{nn}.\label{eq:068}
\end{equation}
Since $\left\langle a^{*}(\omega_{kn})a(\omega_{k'n})\right\rangle =\delta_{kk'}$
because of the statistical independence of the field variables, Eqs.
(\ref{eq:068}) give 
\begin{equation}
C_{n}(xp)=im\sum_{k}\omega_{kn}\left|x_{kn}\right|^{2},\label{eq:070}
\end{equation}
\begin{equation}
C_{n}(px)=-im\sum_{k}\omega_{kn}\left|x_{kn}\right|^{2}.\label{eq:071}
\end{equation}
These are the \emph{ordered }(or\emph{ one-sided})\emph{ covariances}
of $x$ and $p$ and of $p$ and $x$, respectively. Note that
\begin{equation}
\left\langle x^{*}(t)p(t)+p^{*}(t)x(t)\right\rangle _{n}=2x_{nn}p_{nn}\label{eq:072}
\end{equation}
and 
\begin{equation}
\left\langle x^{*}(t)p(t)-p^{*}(t)x(t)\right\rangle _{n}=2im\sum_{k}\omega_{kn}\left|x_{kn}\right|^{2}.\label{eq:073}
\end{equation}
The difference of the covariances is exactly the Poisson bracket in
the $a,a^{*}$-representation, and gives
\begin{equation}
\left\{ x(t),p(t)\right\} _{n}=2im\sum_{k}\omega_{kn}\left|x_{kn}\right|^{2}.\label{eq:074}
\end{equation}
In calculating it we took into account that the left factor must reverse
the effect of the right factor, that the upward (downward) transitions
depend on the normal variable $a^{*}(a)$, and that $\left\{ a,a^{*}\right\} =-\left\{ a^{*},a\right\} $
for all $a_{kn}$.

Comparing the \noun{sed} equations (\ref{eq:072}) and (\ref{eq:073})
with the quantum equations (\ref{eq:065}) and (\ref{eq:066}), we
find that 
\begin{equation}
\left\{ x(t),p(t)\right\} _{n}=\left[\hat{x},\hat{p}\right]_{n},\label{eq:075}
\end{equation}
\begin{equation}
\left\langle x^{*}(t)p(t)+p^{*}(t)x(t)\right\rangle _{n}=\left\{ \hat{x},\hat{p}\right\} _{n}.\label{eq:078}
\end{equation}
Note that in these equations, the left-hand side is the \noun{sed}
expression and the right-hand side is the quantum counterpart. Just
as the commutator $\left[\hat{x},\hat{p}\right]_{n}$ is the Poisson
bracket $\left\{ x(t),p(t)\right\} _{n}$ in the $a,a^{*}$-representation,
the anticommutator $\left\{ \hat{x},\hat{p}\right\} _{n}$ is the
symmetrized covariance of $x(t)$ and $p(t)$. 

In this connection we recall from Section 4 the expression obtained
in the context of \noun{lrt} for the (complex) Fourier transform of
the spectrum $S_{x}(\omega)$, Eq. (\ref{A14b}), containing two terms.
The time-symmetric term is the covariance, which is a measure of the
fluctuations of the system, and the time-asymmetric term is the Poisson
bracket, which contains the response or relaxation function. Classical
\noun{lrt} thus confirms the physical meaning assigned by \noun{sed}
to the quantities used in the quantum description. 

\section{Concluding remarks}
\noindent\begin{flushleft}
Using the tools of \noun{sed}, we have shown that in a stationary
(quantum) state, the particle responds resonantly to the field modes
that can bring it to another stationary state. This response is linear
as long as the field is not very strong, thus allowing an \noun{lrt}
based approach. The response functions are identified with the quantum
matrix elements of the respective dynamical variables, and the canonical
Poisson bracket with the respective quantum commutator. The ordered
covariances, in their turn, are identified with the corresponding
quantum anticommutators. The field variables $a,a^{*}$ to which the
system responds, disappear from the picture, rendering the quantum
description acausal. 
\par\end{flushleft}

Prior to the onset of the quantum regime, the dynamics of the particle,
which is subject to the ever-present random \noun{zpf}, is irreversible.
The quantum formalism gives a coarse-grained (in time, with $\Delta t\sim10^{-17}\,s$)
statistical description of the particle dynamics in the reversible
regime. On this rough timescale, the dynamics appears Markovian, and
indeed the dynamics of quantum processes involving transitions between
quantum states \emph{is} Markovian; see e.g. Ref. \cite{Acca}.

\textbf{Statements and Declarations}

\emph{Competing interests}: The authors have no competing interests.

\emph{Authors' contributions}: Both authors have contributed equally
to this work.

\emph{Data availability}: No data associated in the manuscript.


\begin{thebibliography}{10}
\bibitem{Pava}E. Pavarini, Linear Response Functions, in DMFT at
25: Infinite Dimensions Modeling and Simulation Vol. 4, E. Pavarini,
E. Koch, D. Vollhardt, and A. Lichtenstein, eds., Forschungszentrum
Jülich (2014) ISBN 978-3-89336-953-9 http://www.cond-mat.de/events/correl14

\bibitem{Heis25}W. Heisenberg, Über quantentheoretische Umdeutung
kinematischer und mechanischer Beziehungen. Z. Physik 33(1), 879--893
(1925), DOI:10.1007/bf01328377

\bibitem{foop22}A. M. Cetto, L. de la Peña, Role of the Electromagnetic
Vacuum in the Transition from Classical to Quantum Mechanics, Found.
Phys. 52:84 (2022), doi:10.1007/s10701-022-00605-6

\bibitem{foop24}A. M. Cetto, L. de la Peña, The Radiation Field,
at the Origin of the Quantum Canonical Operators, Found. Phys. 54\textbf{:}51
(2024), doi:10.1007/s10701-024-00775-5

\bibitem{Mars63}T. W. Marshall, Random Electrodynamics, Proc. Royal
Society A: Mathematical, Physical and Engineering Sciences 276, 475--491
(1963), doi:10.1098/rspa.1963.0220

\bibitem{Dice}L. de la Peña, A. M. Cetto, The Quantum Dice. An Introduction
to Stochastic Electrodynamics (Kluwer Academic Publishers, Dordrecht,
1996)

\bibitem{Nernst}W. Nernst, Über einen Versuch, von quantentheoretischen
Betrachtungen zur Annahme stetiger Energieänderungen zurückzukehren,
Verh. Deutsch. Phys. Ges. 18, 83 (1916)

\bibitem{TEQ}L. de la Peña, A. M. Cetto, A. Valdés, The Emerging
Quantum (Springer, Cham, 2015)

\bibitem{vanK}N. G. van Kampen, Stochastic differential equations,
Phys. Rep. 24, 171 (1976)

\bibitem{Pap91}A. Papoulis, Probability, Random Variables, and Stochastic
Processes (McGraw-Hill, Boston, MA, 1991), Chapter 6

\bibitem{Kubo}R. Kubo, Statistical-Mechanical Theory of Irreversible
Processes. I. General Theory and Simple Applications to Magnetic and
Conduction Problems, J. Phys. Soc. Jap. 12 (6), 570--586 (1957).
doi:10.1143/JPSJ.12.570 

\bibitem{Kubobook}R. Kubo, M. Toda, N. Hashitsume, Statistical Physics
II, Springer Series in Solid-State Sciences (Springer, Berlin, 1978)

\bibitem{Nel66}E. Nelson, Derivation of the Schrödinger equation
from Newtonian mechanics, Phys. Rev. 150, 1079 (1966)

\bibitem{Nel12}E. Nelson, Review of stochastic mechanics, J. Phys.
Conf. Ser. 361, 012011 (2012)

\bibitem{Mil}P. W. Milonni, The Quantum Vacuum (Academic Press, 1994)

\bibitem{Wio}H. S. Wio, R. R. Deza, J. M. López, An Introduction
to Stochastic Processes and Nonequilibrium Statistical Mechanics (World
Scientific, NJ, 2012), Chapter 6

\bibitem{ES2022}E. Santos, Realistic Interpretation of Quantum Mechanics
(Cambridge Scholars Publishing, 2022), ISBN 1-5275-7974-3

\bibitem{Acca}L. Accardi, Nonrelativistic QM as a Noncommutative
Markov Process, Adv. Math. 30, 329 (1976).

\end{thebibliography}
\end{document}